\documentclass[12pt]{article}
\usepackage{graphicx}

\usepackage{amsfonts}
\usepackage{amsmath}
\usepackage[english]{babel}
\usepackage{hyperref}
\usepackage{color}
\hypersetup{colorlinks=true,citecolor=blue,linkcolor=blue,filecolor=blue,urlcolor=blue}
\usepackage{tikz}
\usepackage{framed}
\usepackage{setspace}
\usepackage{nicefrac}
\usepackage[shortlabels]{enumitem}

\begin{document}

\title{A QBist reads Merleau-Ponty}

\author{R\"udiger Schack\thanks{Email: r.schack@rhul.ac.uk} \\
{\it Department of Mathematics, Royal Holloway, University of London} \\
{\it Egham, Surrey TW20 0EX, UK}}

\date{1 December 2022}
\maketitle

\begin{abstract}
Following earlier work by Michel Bitbol~\cite{Bitbol2020} and Laura de La Tremblaye~\cite{Tremblaye2020} which examines QBism from the perspective of phenomenology, this short paper explores points of contact between QBism and
Maurice Merleau-Ponty's essay {\it The intertwining---the chiasm}.

\end{abstract}


In his 1814 book {\it A Philosophical Essay on Probabilities}, Pierre-Simon
Laplace wrote the following words: ``An intellect which at a certain moment
would know all forces that set nature in motion, and all {positions} of all
items of which nature is composed, if this intellect were also vast enough to
submit these data to analysis, it would embrace in a single formula the
movements of the greatest bodies of the universe and those of the tiniest atom;
[\ldots] the future just like the past would be present before [this
intellect's] eyes''~\cite[p.~4]{Laplace1902}.

This quotation summarizes the predominant scientific world view in the 19th
century, which is characterized by two main ideas. Firstly, nature is governed
by physical laws that constrain the ``movements of the greatest bodies of the
universe and those of the tiniest atom'', and secondly, it is in principle
possible to have an objective view of the universe from the outside, i.e., from
a god's eye or third-person standpoint.

Quantum mechanics presents a challenge to these ideas. In particular, there is
a strong tension between the third-person view of the universe and the fact
that the observer plays a central role in a quantum measurement. Despite this,
it is probably fair to say that the majority of 21st century scientists
subscribes to the two ideas above. Moreover, the field of quantum foundations
is dominated by efforts to reconcile quantum theory with Laplace's 19th-century
world view. For instance, the many-worlds interpretation of quantum mechanics
postulates the existence of an objective wavefunction of the universe which
evolves in time according to a deterministic law.

Many thinkers have been critical of a third-person account of the world. For
instance, in his recent book {\it Looking East in Winter}, Rowan Williams wrote
``Ironically, an epistemology that knows only the operation of `from
outside' knowledge applied to a material universe is in danger of occluding the
reality of the body in its specific location and embeddedness in a pattern of
interdependence, and of nurturing the fantasy that knowing is essentially the
action of a disembodied subject working on embodied data---the default position
of a good deal of post-medieval Western thinking about knowledge, as if the
object is always somewhere and the subject is nowhere''~\cite[p.~72]{Williams2021}.

Laplace's 19th century world view is simple and tidy, but it runs counter to
one of our most basic experiences as human agents, namely that we possess free
will and are able to choose our actions freely. Philosophy has come up with
many different answers to the question of whether free will can exist in an
objective world governed by physical law. For instance, defenders of {\it
compatibilism\/} claim that free will is compatible with determinism. {\it
Libertarians\/} maintain that the world is indeterministic and that free will
exists. And some authors defend the view, sometimes labeled {\it hard
incompatibilism}, that the laws of physics rule out the notion of free will,
independently of the question of determinism~\cite{OConnor2022}.

Many scientists agree with hard incompatibilism.  For instance, Stephen Hawking
and Leonard Mlodinow wrote: ``It is hard to imagine how free will can operate
if our behaviour is determined by physical law, so it seems that we are no more
than biological machines and that free will is just an illusion''~\cite[p.~32]{Hawking2010}.

The esssential difficulty of reconciling a third-person view of a world
governed by physical law with free will is vividly expressed in this quotation
by Thomas Nagel in his book {\it The View from Nowhere}: ``This naturally
suggests that [\ldots] an account of freedom can be given which is compatible
with the objective view, and perhaps even with determinism. But I believe this
is not the case. All such accounts fail to allay the feeling that, looked at
from far enough outside, agents are helpless and not responsible.
Compatibilist accounts of freedom tend to be even less plausible than
libertarian ones''~\cite[p.~113]{Nagel1989}.

I agree with Nagel here. I thus believe that, in order to preserve freedom, one
has to abandon what Nagel calls the objective view. Fortunately, there exist
alternatives. In this article, I will exhibit points of contact between QBism
and the philosophy of Maurice Merleau-Ponty, both of which explicitly
reject a third-person account of the world.

QBism~\cite{Fuchs2010,Fuchs2016,Fuchs2017,Fuchs2013,Fuchs2014}
resolves the tension between the third-person view of the world and
quantum mechanics through a bold move concerning the nature of
probabilities. QBism maintains that all probabilities, including those equal to
1 or 0 (which express certainty one way or another) and those that are derived
from applying the rules of quantum mechanics, are valuations that an agent
ascribes to his degrees of belief in possible outcomes. In this view,
probabilities are always personal to the agent who assigns them. They get
operational meaning from decision theory: an agent uses probabilities to decide
rationally what action to take in the face of uncertainty.

The personalist, decision-theoretic approach to probability was developed in
the 1920s by Frank Ramsey~\cite{Ramsey1931} and Bruno de Finetti~\cite{deFinetti1931}
and given its modern form by Leonard J.\ Savage~\cite{Savage1954} in the
1950s. This approach to probability is well established in economics and
finance. When used in physics, it is often associated with the idea of
ignorance: when an agent assigns a probability to the outcome of a measurement,
say, of a particle position, the probability expresses the agent's ignorance of
what the outcome will be, but the outcome is determined by the actual particle
position, assumed to exist prior to the measurement. In this reading, a
measurement uncovers a pre-existing property, and agents use probabilities only
because they are ignorant of this property.

QBism's bold move was to retain the personalist nature of probability while at
the same time rejecting the idea that a quantum measurement reveals a
pre-existing property. According to QBism, the world is truly
indeterministic. Neither a measurement outcome nor its probability are given by
physical law. Measurement outcomes are personal consequences for the agent
taking the measurement action and come into existence only through the
measurement action itself. And their probabilities express the agent's
subjective degrees of belief in the possible outcomes.

Probabilities come into the quantum formalism through the Born rule, a
fundamental formula that connects three mathematical objects: a quantum state, a collection of measurement operators, and a probability
distribution. In the usual reading, the quantum state is determined by the
properties of the measured physical system along with the way it was prepared,
while the collection of measurement operators characterizes the measurement
apparatus. Both are regarded as objective quantities. In this reading, the Born
rule thus functions as a physical law that specifies objective outcome
probabilities given quantum state and measurement operators.

By contrast, QBism views the Born rule as a normative constraint on an (or any)
agent's decision making. By calling the Born rule normative, we mean that,
rather than connecting quantities that describe how a part of the world {\it
  is}, it tells the agent how he {\it should\/} act in order to make better
decisions. For a QBist, not only probabilities but also quantum states and
measurement operators express an agent's---typically an experimental
physicist's---degrees of belief. The Born rule now functions as a consistency
criterion which puts constraints on the agent's decision-theoretic beliefs.

To show what this might look like in practice, here are some steps that a
physicist or agent might follow in order to apply the quantum formalism.

Step 1: Identify a part of the world (a ``physical system'') on which to
act. This could be, for instance, an ion in a trap. This step typically
involves an elaborate experimental apparatus enabling the agent to focus on the
precise part of the world he is interested in. QBism considers the apparatus as
an extension of the agent, allowing the agent to access the physical system
directly.

Step 2: Identify a set of measurement outcomes. For a meaningful experiment, an
agent needs to know what he is looking for. In the case of the ion, this could
simply be an answer to the question ``at a given moment, will I see the ion emit
light or not?''. As mentioned earlier, QBism regards the answer as a
consequence of the measurement action which is personal to the agent.

Step 3: Assign probabilities to these outcomes. The probabilities could be
based on, e.g., prior calibration measurements as well as the agent's prior
beliefs.

Step 4: Assign each outcome a measurement operator. This and the next step
requires intimate knowledge of the experimental setup and familiarity with the
quantum formalism. The assignment will depend on the agent's prior beliefs,
which will be shaped by his experience as an experimenter.

Step 5: Assign a quantum state.

Step 6: Check that the assignments in steps 3, 4 and 5 are consistent with the
Born rule. If that is not the case, return to steps 3, 4 and 5 and try to
resolve the inconsistency. There is no unique way of doing this. It could
involve verifying calculations, critically examining assumptions, adjusting the
experimental setup, making additional calibration measurements, etc. The
normative content of the Born rule is that it requires agents to strive for
consistency, without providing them with instructions as to how to get there.

According to QBism, quantum mechanics is a small theory. It does not attempt to
describe the inner workings of physical systems, let alone the universe. It
only provides agents with a criterion of consistency to strive for. But it is
a small theory with enormous power. By applying its formalism consistently,
physicists and engineers have created modern technology.

Since agents and users of quantum mechanics take center stage in QBism, it will
be useful to say explicitly what we mean by these terms: {\it Agents\/} are
entities that (i) can take actions freely on parts of the world external to
themselves so that (ii) the consequences of their actions matter for them. And
a {\it user of quantum mechanics\/} is an agent that is capable of
applying the quantum formalism normatively. These definitions make sure that
it is meaningful for an agent to use decision theory without restricting the
class of agents too much.

To briefly summarize our discussion of QBism so far, we have seen that QBism
regards quantum mechanics as a normative tool for agents to make better
decisions. Probabilities are personal to each agent and are not given by
physical law. Quantum measurement outcomes are personal to the agent making the
measurement and do not pre-exist the measurement. The world does not admit a
third-person description and is fundamentally indeterministic. Agents possess
genuine freedom.

QBism's spirit is beautifully captured in this quotation by William James: ``Why
may not the world be a sort of republican banquet [\ldots], where all the
qualities of being respect one another's personal sacredness, yet sit at the
common table of space and time?  To me this view seems deeply probable. Things
cohere, but the act of cohesion itself implies but few conditions, and leaves
the rest of their qualifications indeterminate. [\ldots] [I]f we stipulate only
a partial community of partially independent powers, we see perfectly why no
one part controls the whole view, but each detail must come and be actually
given, before, in any special sense, it can be said to be determined at
all. This is the moral view, the view that gives to other powers the same
freedom it would have itself [\ldots]''~\cite{James1956}.

What remains a big challenge is to develop an explicit ontology for QBism, an
ontology that agrees with the spirit of William James's ``republican banquet'',
i.e., an ontology based on first-person experience but which accounts for a
real world beyond any particular agent's experience and thus avoids the trap of
idealism. It turns out that there is a significant overlap between the project
of finding such a QBist ontology and the philosophy of Maurice Merleau-Ponty
and other phenomenologists. Below I will sketch some points of contact between
QBism and Merleau-Ponty's essay {\it The intertwining---the chiasm\/}~\cite{Merleau-Ponty1968}.

In that essay, Merleau-Ponty asks three main questions: What does it mean to
experience the world through seeing and touching? What is the nature of our
relationship with others? What is the status of our ideas about the world? He
starts out with an implicit acknowledgement of the great difficulty of
answering these questions in a philosophical framework that takes first-person
experience as its starting point: ``once again [philosophy] must recommmence
everything,'' he writes~\cite[p.~130]{Merleau-Ponty1968}. Like QBism, he rejects both
idealism and a third-person view of the world: ``We have to reject the age-old
assumptions that put the body in the world and the seer in body, or,
conversely, the world and the body in the seer as in a box''~~\cite[p.~138]{Merleau-Ponty1968}.

QBism does not only rule out a third-person view of the whole world, it also
rejects the notion that a quantum state can provide a complete description of any
particular physical system. Firstly, an agent's quantum state for a
physical system is not a description of the system at all, but an encoding of an
agent's expectations regarding the consequences of his actions on the
system. Secondly, before writing down a quantum state, the agent will have to
decide what (necessarily) limited set of aspects to focus on, i.e., identify a
set of potential measurement outcomes. For instance, this could be the range of
wavelengths that a certain spectrometer is able to resolve. The outcome of the
measurement will be personal to the agent and will be the answer to a very
specific question. It will never capture all there is to say about the
system.

Merleau-Ponty expresses the idea that an agent's experience is always much
richer than any label one might put on it in a striking passage: ``[\ldots]
this red under my eyes is not, as is always said, a quale, a pellicle of being
without thickness, a message at the same time indecipherable and evident, which
one has or has not received, but of which, if one has received it, one knows
all there is to know, and of which in the end there is nothing to say.''
Rather, it is a boundlessly rich experience, a ``punctuation in the field of
red things, which includes the tiles of roof tops, the flags of gatekeepers and
of the Revolution, certain terrains near Aix or in Madagascar, it is also a
punctuation in the field of red garments, which includes, along with the
dresses of women, robes of professors, bishops, and advocate generals, and also
in the field of adornments and that of uniforms''~\cite[p.~132]{Merleau-Ponty1968}.

Merleau-Ponty's answer to the question of ontology is the concept of ``flesh'',
which he compares to the elements of ancient Greek philosophy, water, air,
earth, and fire: ``[flesh] is an element of being''~\cite[p.~139]{Merleau-Ponty1968}.
It is flesh which sustains the rich experience of colours and other
visibles. He writes that ``[flesh] is the tissue that lines them, sustains
them, nourishes them, and which for its part is not a thing, but a possibility,
a latency [\ldots] flesh is not matter, is not mind, is not substance [\ldots]
it is a general thing, midway between the spatio-temporal individual and the
idea, a sort of incarnate principle that brings a style of being wherever there
is a fragment of being''~\cite[p.~139]{Merleau-Ponty1968}.

A key idea of QBism is that there is no such thing as passive observation of a
quantum system. Any measurement is an active intervention. To experience a
response from a system, an agent needs to touch the system, typically through a
measurement apparatus which in QBism should be understood as an extension of
the agent himself. This is why QBists tend to talk about measurement {\it
  actions}, rather than just measurements, and this is why QBists rarely use
the term ``observer,'' but prefer the term ``agent'' instead.

The touching subject plays a central role in Merleau-Ponty's philosophy. He
uses the example where ``my right hand touches my left hand while it is
palpating the things'' to show that the touching subject is part of the
palpable world itself, ``such that the touch is formed in the midst of the
world and as it were in the things''~\cite[p.~134]{Merleau-Ponty1968}. He argues
that an analagous relation exists between the seeing subject and the visible,
i.e., that a seeing subject is necessarily part of the visible world. I can
touch and see things precisely because I am visible and tangible myself,
because the visible is an archetype for the seeing and vice versa. Ultimately
it is flesh---a relation of the visible with itself that ``constitutes me as a
seer''~\cite[p.~140]{Merleau-Ponty1968}. Again quoting Rowan Williams, ``In the
striking phrase used by Orion Edgar in his study of the theological
implications of Merleau-Ponty's philosophy, `nature lies on both sides of
perception'\,''~\cite[p.~73]{Williams2021}.

In QBism, the idea that touching and being touched, or seeing and being seen,
are two sides of the same coin is expressed through this ``Copernican
principle'': ``By one category of thought we are agents, but by
another category of thought we are physical systems. And when we take actions
upon each other, the category distinctions are symmetrical''~\cite{Fuchs2013}.

That there are other agents is one of the most fundamental human experiences,
and both QBism and Merleau-Ponty acknowledge this explicitly. According to
QBism, one agent can apply the quantum formalism to another agent in the same
way as to any physical system. As for any physical system, this allows the
agent to bet on the consequences of his actions on the other agent. And as for
any physical system, this leaves the fundamental autonomy of the other agent
intact. To repeat William James's words quoted above, this is ``the moral view,
the view that gives to other powers the same freedom as it would have itself
[\ldots]''.

For Merleau-Ponty it is again flesh which provides a connection between the
first-person experiences of different subjects. To paraphrase him, this is
possible as flesh is unversal, not particular to me and my body and my
experience of the world.  Hence it enables the synergy not only between
different organs in my body, e.g., my hands and eyes, or between my body and
the visible and palpable world around me, but also between myself and
others. This happens ``[\ldots] by virtue of the fundamental fission or
segregation of the sentient and the sensible which, laterally, makes the organs
of my body communicate and founds transitivity from one body to another''~\cite[p.~143]{Merleau-Ponty1968}.

Merleau-Ponty contrasts his vision of a world that has inexhaustible depth and
contains first-person experiences other than our own with the ``solipsistic
illusion'' that an agent might capture from his perspective all there is to say
about the world: ``But what is proper to the visible is, we said, to be the
surface of an inexhaustible depth: this is what makes it able to be open to
visions other than our own. In being realized, they therefore bring out the
limits of our factual vision, they betray the solipsist illusion that consists
in thinking that every going beyond is a surpassing accomplished by oneself''~\cite[p.~143]{Merleau-Ponty1968}.  Other agents are fundamental to Merleau-Ponty's
philosophy: ``[\ldots] for the first time, through the other body, I see that,
in its coupling with the flesh of the world, the body contributes more than it
receives, adding to the world that I see the treasure necessary for what the
other body sees''~\cite[p.~144]{Merleau-Ponty1968}.

When an experimental physicist selects a physical system to act on, indentifies
a sample space of potential measurement outcomes, chooses a measurement
apparatus, assigns probabilities, quantum states and measurement operators and
finally takes a measurement action, this is when Merleau-Ponty's ``fundamental
fission or segregation of the sentient and the sensible'' occurs. QBism regards
the measurement apparatus as an extension of the agent. Translated into
Merleau-Ponty's language, the measurement apparatus thus becomes a part of the
``sentient'', whereas the ``sensible'' (i.e., the palpable or visible)
corresponds to the physical system that the agent is acting on.

For example, if I measure a qubit, the bit (0 or 1) I obtain as my measurement
outcome is just a tiny aspect of the particle in front of me.  What allows me
to focus on this tiny aspect and ignore everything else about the particle is
the measurement apparatus.  Equipped with the apparatus as an extension of
myself, the measurement can be seen as me touching the qubit and experiencing
the outcome directly. If I understand Merleau-Ponty correctly, my touching of
the qubit is an instance of the ``fission of the sentient and the sensible''.

Coming back to the question of other agents, QBism's Copernican principle has a
compelling application to the resolution~\cite{DeBrota2020} of the
famous Wigner's friend paradox~\cite{Wigner1961}. In a nutshell, the paradox arises
from a thought experiment in which Wigner applies the quantum formalism to a
lab containing his friend. Since Wigner is outside the lab and has assigned a
quantum state to the total system including the friend, it appears as if Wigner
were in a privileged position vis-a-vis his friend. This leads to the
paradox. It arises from a failure to treat the friend as an autonomous agent on
the same footing as Wigner himself, in violation of the symmetry required by
the QBist Copernican principle.

For instance, in their recent version of the paradox, Baumann and Brukner~\cite{Baumann2020} argue
that the friend should base her predictions on Wigner's state assignment as well
as her personal experience. But if the friend is an autonomous QBist agent, in
order to apply the quantum formalism consistently, she must treat Wigner and
the relevant parts of the laboratory surrounding her as a physical system
external to herself, just as Wigner treats her and the laboratory as a physical
system external to himself. It does not matter that the laboratory spatially
surrounds the friend; it, like the rest of the universe, is external to her
agency, and that is what counts. Once symmetry is restored in this
way, the paradox disappears.

The root of the paradox is the notion that Wigner's state assignment is able to
capture everything that is relevant to the friend's experience. One could argue
that this turns the Wigner of the thought experiment into a solipsist. In
Merleau-Ponty's words, he has fallen for ``the solipsist illusion that consists
in thinking that every going beyond is a surpassing accomplished by
oneself''. QBist Wigner, by contrast and quoting William James's republican
banquet passage, ``gives to [his friend] the same freedom [he] would have
[himself]''.

To conclude, there exist two radically different scientific world views. On the
one hand, there is the mainstream third-person perspective of a world which
evolves according to laws and in which agents are biological machines. On the
other hand, there is the QBist view which rejects a third-person perspective of
the world, in which agents matter fundamentally, and where laws have a
normative character. It is obvious which of the two views a free agent
should adopt.

This work was supported by Grant 62424 from the John Templeton Foundation. The
opinions expressed in this publication are those of the author and do not
necessarily reflect the views of the John Templeton Foundation.

\end{document}